\documentclass[letter]{jpsj2}

\begin{document}

\def\runtitle{
Universal Ratio of Seebeck Coefficient and Specific Heat
}
\def\runauthor{Kazumasa {\sc Miyake} and Hiroshi {\sc Kohno}
}

\title
{Theory of Quasi-Universal Ratio of Seebeck Coefficient to Specific Heat 
in Zero-Temperature Limit in Correlated Metals}

\author{Kazumasa {\sc Miyake} and Hiroshi {\sc Kohno}} 

\inst
{
Division of Materials Physics, Department of Materials Engineering Science, 
Graduate School of Engineering Science \\
Osaka University, Toyonaka 560-8531\\
}

\recdate
{August 30, 2004
}

\abst
{
It is shown that the quasi-universal ratio $q=\lim_{T\to0}eS/C\sim\pm1$ 
of the Seebeck coefficient to the specific heat in the limit of $T=0$ 
observed in a series of strongly correlated metals can be understood on the 
basis of the Fermi liquid theory description.  
In deriving this result, it is crucial that a relevant scattering arises 
from impurities, but not from the mutual scattering of quasiparticles.  
The systematics 
of the sign of $q$ is shown to reflect the sign of the logarithmic derivative 
of the density of states and the inverse mass tensor of the quasiparticles, 
explaining the systematics of experiments.  In particular, 
the positive sign of $q$ for Ce-based and $f^{3}$-based heavy fermions, and 
the negative sign for Yb-based and $f^{2}$-based heavy fermions, are 
explained.  The case of non-Fermi liquid near the quantum critical point 
(QCP) is briefly mentioned, showing that the ratio $q$ decreases considerably 
toward antiferromagnetic QCP while it remains essentially unchanged for the 
ferromagnetic QCP or QCP due to a local criticality.  

}

\kword
{Seebeck coefficient, specific heat, Fermi liquid, strongly correlated metals, 
periodic Anderson model
}

\sloppy
\maketitle
Strongly correlated electron systems exhibit some quasi-universal ratios among 
various physical quantities.  The Wilson ratio~\cite{Wilson}, $\chi/\gamma$, 
of the magnetic susceptibility $\chi$ to the Sommerfeld 
coefficient $\gamma$, and the so-called Kadowaki-Woods ratio~\cite{KW} 
$A/\gamma^{2}$, $A$ being the coefficient of the $T^{2}$-term of the 
resistivity, 
take a quasi-universal value in a series of strongly correlated electron 
systems up to logarithmic accuracy.  The validity of the Fermi liquid theory 
and the dynamical nature of the mass enhancement lie behind the 
universality~\cite{MMV,Yamada}. Quite recently, Behnia, Jaccard, and Flouquet 
have revealed that the ratio of the Seebeck coefficient $S$ 
to the specific heat $\gamma T$ takes a quasi-universal value, in the 
low-temperature limit, for a series of correlated compounds~\cite{Behnia}.  
The purpose of this Letter is to clarify the reason for the universal 
value for $eS/\gamma T$ on the basis of the Fermi liquid theory for 
thermoelectricity~\cite{Kohno,Kontani}.  

A firm starting point is that the canonical formula for the Seebeck 
coefficient $S$ in the zero-temperature limit~\cite{Ziman} is valid even 
though strong correlations are apparent among the electrons:
\begin{equation}
S=-{\pi^{2}\over 3}{k_{\rm B}^{2}T\over e}\left({\partial\ln\sigma(\epsilon)
\over\partial\epsilon}\right)_{\epsilon=\mu},
\label{eq:1}
\end{equation}
where $e$ ($>0$) is the elementary charge, $\sigma(\epsilon)$ is the 
conductivity of electrons with energy 
$\epsilon$, and $\mu$ is the chemical potential at $T=0$.  Although the above 
expression was derived for a single-band Fermi liquid~\cite{Kohno,Kontani}, 
there is no reason for doubting its validity also in the multiband 
Fermi liquid~\cite{Phonon1}.  
Therefore, we discuss the Seebeck coefficient on the basis of formula 
(\ref{eq:1}).  Then, we use the conductivity for a multicarrier system: 
\begin{equation}
\sigma(\epsilon)=\sum_{i}\sigma_{i}(\epsilon),
\label{eq:2}
\end{equation}
in the expression of the Seebeck coefficient, (\ref{eq:1}).  Here, 
$\sigma_{i}$ denotes the conductivity due to the $i$-th Fermi surface of 
multicomponent carriers.  By some algebra, 
we obtain the rule of summing the Seebeck coefficient of each component as 
follows: 
\begin{equation}
S=\sum_{i}{\sigma_{i}\over\sigma}S_{i}.
\label{eq:4}
\end{equation}

\subsection*{Single-Band Fermi liquid}
First, we discuss the case of heavy fermions with a single Fermi surface.  
The conductivity of a single-band Fermi liquid is approximated as~\cite{Yamada}
\begin{equation}
\sigma(\epsilon)=e^{2}\tau(\epsilon)\sum_{{\vec k}}
\delta(\epsilon-\epsilon({\vec k}))
\langle v_{\alpha}({\vec k})v^{*}_{\alpha}({\vec k})\rangle, 
\label{eq:5}
\end{equation}
where $v^{*}$ denotes the renormalized velocity by vertex corrections due to 
the many-body effect, and $\langle\cdots\rangle$ denotes the average over 
the direction $\alpha$\ ($=x,y,z$).  
Here, ${\vec v}({\vec k})\equiv\partial E_{{\vec k}}/\partial {\vec k}$, 
$E_{{\vec k}}$ being the dispersion of the quasiparticle, and $\tau(\epsilon)$ 
denotes the transport collision time of the quasiparticles with energy 
$\epsilon$.  Hereafter, the 
renormalized velocity $v^{*}_{\alpha}$ is approximated by $v_{\alpha}$, 
because the impurity scattering plays a dominant role for which such 
renormalization is not crucial.  

The transport collision time for the Fermi liquid is given, around the Fermi 
level in the zero-temperature limit, as 
\begin{equation}
\tau(\epsilon)^{-1}=\tau_{\rm imp}(\epsilon)^{-1}+B(\epsilon-\mu)^{2},
\label{eq:6}
\end{equation}
where $\tau_{\rm imp}(\epsilon)$ is the collision time due to impurity 
scattering, and $B$ is a constant of the order of the inverse effective 
Fermi energy.  
The second term in the collision rate (\ref{eq:6}) makes no contribution to 
$S$ since the following relation holds:
\begin{equation}
\lim_{\epsilon\to\mu}{\partial\over\partial\epsilon}\ln\tau(\epsilon)
=\lim_{\epsilon\to\mu}{\partial\over\partial\epsilon}\ln
\tau_{\rm imp}(\epsilon).
\label{eq:7}
\end{equation}
Therefore, the Seebeck coefficient for a single-band Fermi liquid 
is expressed as 
\begin{eqnarray}
S&=&-{\pi^{2}\over 3}{k_{\rm B}^{2}T\over e}
\lim_{\epsilon=\mu}\Bigg\{
{\partial\over\partial\epsilon}\ln\tau_{\rm imp}(\epsilon)
\nonumber
\\
& &\qquad+{\sum_{{\vec k}}\delta(\epsilon-E_{{\vec k}})
\langle\partial v_{\alpha}({\vec k})/\partial k_{\alpha}\rangle\over
\sum_{{\vec k}}\delta(\epsilon-E_{{\vec k}})
\langle v_{\alpha}({\vec k})v_{\alpha}({\vec k})\rangle}
\Bigg\},
\nonumber
\\
&\equiv&-{\pi^{2}\over 3}{k_{\rm B}^{2}T\over e}{\cal F}.
\label{eq:8a}
\end{eqnarray}

The collision rate due to impurity scattering in the $t$-matrix approximation 
is given by
\begin{equation}
\tau_{\rm imp}(\epsilon)^{-1}={2\pi z^{2}(\epsilon)N^{*}(\epsilon)u^{2}
\over 1+[\pi z(\epsilon)N^{*}(\epsilon)u]^{2}}\times c_{\rm imp}, 
\label{eq:9}
\end{equation}
where $c_{\rm imp}$ is the impurity concentration, $N^{*}(\epsilon)$ is 
the density of states (DOS) of the renormalized quasiparticles, 
and $z(\epsilon)$ is the renormalization amplitude~\cite{Doniach}.  
In deriving eq.(\ref{eq:9}), we have assumed the $s$-wave impurity potential 
and also neglected the real part of the electron self-energy due to impurity 
scattering.  Then, the logarithmic derivative of $\tau_{\rm imp}$ is given as
\begin{equation}
\lim_{\epsilon\to\mu}{\partial\over\partial\epsilon}\ln
\tau_{\rm imp}(\epsilon)=
{[\pi z(\mu)N^{*}(\mu)u]^{2}-1\over
[\pi z(\mu)N^{*}(\mu)u]^{2}+1}\times
{\partial\over \partial\epsilon}
\ln N^{*}(\epsilon)\biggl|_{\epsilon=\mu}.  
\label{eq:10}
\end{equation}
Here, we have used the estimation 
$\lim_{\epsilon\to\mu}|\partial \ln z(\epsilon)/\partial\epsilon|
\ll\lim_{\epsilon\to\mu}|\partial \ln N^{*}(\epsilon)/\partial\epsilon|$, 
which has been verified by explicit calculations, {\it e.g.}, FLEX 
approximation for $t$-$t'$ Hubbard model near the 
half-filling~\cite{Hoshihara,Phonon2}.  
It is noted that $z(\mu)N^{*}(\mu)$ is of the 
same order as the unrenormalized one since the mass enhancement in $N^{*}$ is 
cancelled by the smallness of $z$, while the logarithmic derivative 
of $N^{*}$ is highly enhanced in a heavy fermion situation.  

In order to make the discussion explicit, we proceed with the periodic 
Anderson model (PAM):
\begin{eqnarray}
H_{\rm PAM}&=&\sum_{{\vec k}}\left\{\epsilon_{{\vec k}}
c^{+}_{{\vec k}\sigma}c_{{\vec k}\sigma}+
\varepsilon_{\rm f}f^{+}_{{\vec k}\sigma}f_{{\vec k}\sigma}
+\left[V_{{\vec k}}c^{+}_{{\vec k}\sigma}f_{{\vec k}}+{\rm h.c.}\right]\right\}
\nonumber
\\
& &\qquad\quad
+U\sum_{i}n^{\rm f}_{i\uparrow}n^{\rm f}_{i\downarrow},
\label{eq:11}
\end{eqnarray}
where the notations are conventional.  According to the Fermi liquid theory 
based on PAM~\cite{Yamada}, the effective Hamiltonian of quasiparticles 
near the Fermi level is given as 
\begin{equation}
H_{\rm qp}=\sum_{{\vec k}}\left\{\epsilon_{{\vec k}}
c^{+}_{{\vec k}\sigma}c_{{\vec k}\sigma}+
{\tilde \varepsilon}_{\rm f}f^{+}_{{\vec k}\sigma}f_{{\vec k}\sigma}
+\left[{\tilde V}_{{\vec k}}c^{+}_{{\vec k}\sigma}f_{{\vec k}}
+{\rm h.c.}\right]\right\},
\label{eq:12}
\end{equation}
where the renormalized $f$-level is determined by the $f$-electron self-energy 
$\Sigma_{\rm f}$ and the renormalization amplitude $a_{\rm f}$ as, 
\begin{equation}
{\tilde \varepsilon}_{\rm f}\equiv
a_{\rm f}[\varepsilon_{\rm f}+\Sigma_{\rm f}(\mu)],
\label{eq:12a}
\end{equation} 
and ${\tilde V}_{{\vec k}}=\sqrt{a_{\rm f}}V_{{\vec k}}$.  
The $k$-dependence of $\Sigma_{\rm f}$ should be rather weaker than 
$N_{\rm F}^{\rm cond}|V|^{2}$, because it is the only way to realize the 
heavy fermion state by satisfying the two conditions, the Landau-Luttinger sum 
rule~\cite{Luttinger,AGD} and the requirement that the $f$-electron 
number $n_{\rm f}$ at each site be nearly unity, $n_{\rm f}\simeq 1$.  
Then, the dispersion of the quasiparticles is given as 
\begin{equation}
E_{\vec k}={1\over 2}\left[
\epsilon_{{\vec k}}+{\tilde \varepsilon}_{\rm f}
-\sqrt{(\epsilon_{{\vec k}}-{\tilde \varepsilon}_{\rm f})^{2}
+4|{\tilde V}_{{\vec k}}|^{2}}\right],
\label{eq:13}
\end{equation}
where we have assumed the Fermi level is located at the bonding band which 
is considered to correspond to the Ce-based heavy fermions.  

Then, the DOS of the quasiparticles, 
$N^{*}(\epsilon)\equiv2\sum_{{\vec k}}\delta(\epsilon-E_{\vec k})$, 
is given by the standard calculation as follows: 
\begin{equation}
N^{*}(\epsilon)=2N^{\rm cond}
{\epsilon^{*}(\epsilon)-2\epsilon+{\tilde \varepsilon}_{\rm f}
\over {\tilde \varepsilon}_{\rm f}-\epsilon},
\label{eq:14}
\end{equation}
where $N^{\rm cond}$ is the DOS of the conduction electrons, and 
$\epsilon^{*}(\epsilon)$ is determined by the following equation: 
\begin{equation}
\epsilon={1\over 2}\left[
\epsilon^{*}(\epsilon)+{\tilde \varepsilon}_{\rm f}
-\sqrt{(\epsilon^{*}(\epsilon)-{\tilde \varepsilon}_{\rm f})^{2}
+4|{\tilde V}_{{\vec k}}|^{2}}\right], 
\label{eq:15}
\end{equation}
and is of the order of $D$, half the bandwidth of conduction electrons.  
Since $|\epsilon^{*}(\epsilon)-\mu|\gg
|{\tilde \varepsilon}_{\rm f}-\mu|$, the DOS is given near the Fermi level, 
$\epsilon=\mu$, as 
\begin{equation}
N^{*}(\epsilon)\sim 2N^{\rm cond}
{D\over |{\tilde \varepsilon}_{\rm f}-\epsilon|}
={b\over |{\tilde \varepsilon}_{\rm f}-\epsilon|}, 
\label{eq:16}
\end{equation}
where $b$ is a constant of ${\cal O}(1)$ because 
$N^{\rm cond}D\sim1$.  Thus, the logarithmic derivative of DOS is 
given by the following simple expression: 
\begin{equation}
{\partial\over\partial\epsilon}\ln N^{*}(\epsilon)\biggl|_{\epsilon=\mu}=
-{1\over \mu-{\tilde \varepsilon}_{\rm f}}.
\label{eq:17}
\end{equation}
It is noted that the denominator is of the order of the effective Fermi 
energy.  

Now we discuss the second term in the brace of (\ref{eq:8a}).  With the use of 
the dispersion relation (\ref{eq:13}), the velocity of quasiparticles 
${\vec v}_{\alpha}({\vec k})\equiv \partial E_{{\vec k}}/\partial k_{\alpha}$ 
is estimated near the Fermi level as
\begin{equation}
{\vec v}_{\alpha}({\vec k})
={{\tilde \varepsilon}_{\rm f}-E_{{\vec k}}\over
{\tilde \varepsilon}_{\rm f}+\epsilon_{{\vec k}}-2E_{{\vec k}}}
{\partial\epsilon_{{\vec k}}\over\partial k_{\alpha}}
\simeq{{\tilde \varepsilon}_{\rm f}-\mu\over\epsilon^{*}(\mu)}
{\partial\epsilon_{{\vec k}}\over\partial k_{\alpha}},
\label{eq:18}
\end{equation}
where we have considered the fact 
$|{\tilde \epsilon}_{\rm f}-E_{{\vec k}}|\ll \epsilon^{*}(\mu)$.  
The inverse mass tensor is given as 
\begin{equation}
{\partial v_{\alpha}({\vec k})\over\partial k_{\alpha}}
\simeq
-2{{\tilde \varepsilon}_{\rm f}-E_{{\vec k}}\over\epsilon^{*}(\mu)^{2}}
\left({\partial\epsilon_{{\vec k}}\over\partial k_{\alpha}}\right)^{2}
+{{\tilde \varepsilon}_{\rm f}-E_{{\vec k}}\over\epsilon^{*}(\mu)}
{\partial^{2}\epsilon_{{\vec k}}\over\partial k_{\alpha}^{2}}. 
\label{eq:19}
\end{equation}
Therefore, the second term in the brace of (\ref{eq:8a}) is calculated as 
\begin{equation}
{\sum_{{\vec k}}\delta(\epsilon-E_{{\vec k}})
\partial v_{\alpha}({\vec k})/\partial k_{\alpha}\over
\sum_{{\vec k}}\delta(\epsilon-E_{{\vec k}})
v_{\alpha}({\vec k})v_{\alpha}({\vec k})}
={2-\eta\over \epsilon-{\tilde \varepsilon}_{\rm f}},
\label{eq:20}
\end{equation}
where $\eta\equiv\epsilon^{*}(\mu)
(\partial^{2}\epsilon_{{\vec k}}/\partial k_{\alpha}^{2})/
(\partial\epsilon_{{\vec k}}/\partial k_{\alpha})^{2}$, which is equal to 
1/2 for the free electron dispersion and 0 for the linear dispersion of 
conduction electrons.  

Collecting the relations (\ref{eq:10}), (\ref{eq:17}), and (\ref{eq:20}), 
the factor ${\cal F}$ in eq.(\ref{eq:8a}) is expressed as 
\begin{equation}
{\cal F}={3-\eta+(1-\eta)[\pi z(\mu)N^{*}(\mu)u]^{2}
\over 1+[\pi z(\mu)N^{*}(\mu)u]^{2}}\times
{1\over \mu-{\tilde \varepsilon}_{\rm f}}.
\label{eq:21}
\end{equation}
Therefore, the ratio $q$ of the Seebeck coefficient given by eq.(\ref{eq:8a}) 
to the specific heat $\gamma T$ is reduced to the concise form of 
\begin{equation}
q\equiv{S\over T}{e\over \gamma}=
{3-\eta+(1-\eta)[\pi z(\mu)N^{*}(\mu)u]^{2}\over 
1+[\pi z(\mu)N^{*}(\mu)u]^{2}}
\times{1\over b},
\label{eq:23}
\end{equation}
where $\gamma\equiv\pi^{2}k_{\rm B}^{2}N^{*}(\mu)/3$ is the Sommerfeld 
coefficient, and $b\equiv N^{*}(\mu)({\tilde \varepsilon}_{\rm f}-\mu)$.  
In deriving eq.(\ref{eq:23}), we have used eq.(\ref{eq:16}).  
The ratio $q$ depends on the character of the impurity 
scattering:  For a weak impurity potential producing the Born scattering 
($zN^{*}u\ll 1$), $q\simeq (3-\eta)/b$, while $q\simeq (1-\eta)/b$ for the 
impurity with the unitarity scattering ($zN^{*}u\gg 1$).  
The strength of scattering by actual impurities extends from weak to 
strong coupling.  Therefore, we should have taken an average over the 
impurities when we derived relation (\ref{eq:9}), rather than simply 
multiplying the impurity concentration $c_{\rm imp}$.  If we do so, however, 
a simple expression as (\ref{eq:23}) will not be obtained.  
Therefore, for simplicity, we simply take the average of expression 
(\ref{eq:23}) over the strength of the impurity potential.  As expected 
physically, most impurities will have an intermediate character.  
Therefore, it is reasonable to 
expect that the ratio is of the order of ${\cal O}(1)$, leading to 
the quasi-universal value for $q\sim +1$ for Ce-based heavy fermions.  
A crucial point here is that strong renormalization effects in $S$ and 
$\gamma$ cancel each other out, as in the  case of the Kadowaki-Woods 
ratio~\cite{MMV,Yamada}.  

In the case of Yb-based heavy fermions, we should apply the hole version 
of the above discussions, in which the Fermi level is located on the 
antibonding band in an electron picture.  Even in this case, expressions 
(\ref{eq:14}), and (\ref{eq:16})-(\ref{eq:21}), are valid.  
Since $\mu>{\tilde \varepsilon}_{\rm f}$, the sign of (\ref{eq:23}) changes, 
leading to $q\sim -1$.  These results explain the observed 
universal behavior of the Yb-based heavy fermion compounds~\cite{Behnia}.   

\subsection*{Non-Fermi Liquid near Quantum Critical Point}
In the case near the antiferromagnetic (AF) quantum critical point (QCP), 
where the quasiparticles are still well defined, the DOS $N^{*}(\epsilon)$ is 
nearly symmetric around the chemical potential.  Indeed, near the AF-QCP 
in 3d, 
$N^{*}(\epsilon)\propto\{{\rm const.}-
[(\epsilon-\mu)^{2}+\omega_{0}^{2}]^{1/4}\}
\times N^{*}_{\rm loc}(\epsilon)$, 
with $\omega_{0}$ being the energy scale inversely 
proportional to the staggered susceptibility and $N^{*}_{\rm loc}(\epsilon)$ 
being the DOS renormalized by a local correlation effect as discussed above.  
It is noted that $\omega_{0}$ has $T$-dependence, $\propto T^{3/2}$, 
in general.  
Therefore, $\lim_{\epsilon\to\mu}\partial\ln N^{*}(\epsilon)/\partial\epsilon
=\lim_{\epsilon\to\mu}\partial\ln N^{*}_{\rm loc}(\epsilon)
/\partial\epsilon$.  
Furthermore, the second term in eq.(\ref{eq:8a}) is shown to remain unenhanced 
by the AF critical fluctuations.  Indeed, the velocity $v_{\alpha}$ appearing 
in eq.(\ref{eq:20}) depends crucially on the position on the Fermi surface, 
and 
vanishes in proportion to the inverse of the logarithm of its energy on the 
{\it hot lines} but remains finite otherwise.  
Then, the summation with respect to ${\vec k}$ on the 
Fermi surface remains of the same order of magnitude as away from the 
QCP~\cite{comment2}.  

Therefore, the ratio $q$ near the AF-QCP is given as 
$|q|\sim N^{*}_{\rm loc}(\epsilon)/N^{*}(\epsilon)|_{\epsilon=\mu}$, 
which is far less than unity at AF-QCP because $N^{*}(\mu)$ is enhanced 
compared to $N_{\rm loc}(\mu)$ by AF critical fluctuations, although it 
does not diverge. On the other hand, near the ferromagnetic (F) QCP, $q$ 
remains of the same order as that away from QCP, because 
the second term of eq.(\ref{eq:8a}) is proportional to the mass enhancement 
factor both due to local correlation, given by eq.(\ref{eq:20}), and 
the F critical fluctuations which enhance the effective mass 
equally at all points on the Fermi surface, in contrast to the AF critical 
fluctuations discussed above.  The estimation of the first term in 
eq.(\ref{eq:8a}) is also valid for F-QCP.  Then, the ratio $q$ exhibits only a 
small decrease leaving $q\sim\pm 1$ with a logarithmic accuracy.  

Thus, the ratio $q$ considerably decreases 
toward the AF-QCP but does not change appreciably for the F-QCP.  
The estimation for F-QCP is also valid for the local quantum criticality, 
where all the points on the Fermi surface are subject to the effect of 
critical fluctuations~\cite{Si,Holmes}.  
These predictions may be explored by experiments 
around QCP tuned by altering the pressure or the magnetic field.  

\subsection*{Multiband Fermi liquid}
Actual heavy fermion compounds have a multiband structure and plural Fermi 
surfaces.  Therefore, one should calculate $S_{i}$, the contribution from 
the $i$-th band, and sum up by using formula (\ref{eq:4}) with weight 
$\sigma_{i}/\sigma$.  In the zero-temperature limit, where 
only the impurity scattering is relevant to the thermoelectricity, 
$\sigma_{i}/\sigma$ is of the same order for all of the bands because the 
renormalization effect cancels out for the impurity scattering~\cite{Langer} 
unless a renormalization of the impurity potential due to the quantum critical 
fluctuations develops~\cite{MN,MM}.  
Indeed, the conductivity of light ($\ell$) and heavy (h) bands is given by 
\begin{equation}
\sigma^{\ell,{\rm h}}\sim e^{2}\tau_{\rm imp}(v_{\rm F}^{\ell,{\rm h}})^{2}
N_{\rm F}^{\ell,{\rm h}},
\label{eq:24}
\end{equation}
where the collision time $\tau_{\rm imp}$ is given by 
\begin{equation}
(\tau_{\rm imp}^{\ell,{\rm h}})^{-1}=
2\pi^{2}c_{\rm imp}u^{2}(z^{\ell,{\rm h}})^{2}N_{\rm F}^{\ell,{\rm h}}.  
\label{eq:25}
\end{equation}
It is easy to see that the renormalization factors $z$ are cancelled among 
those included in $v\propto z$ and $N_{\rm F}\propto z^{-1}$.  
Then, $\sigma^{\ell}\sim\sigma^{\rm h}$ if the impurity potentials 
for light and heavy carriers are comparable as expected~\cite{Yamada2}.  
Therefore, the contribution of $S_{i}$ from the heavier band dominates 
because $|S^{\rm h}|\gg|S^{\ell}|$.  

Some heavy fermion compounds such as CeCu$_2$Si$_2$ are compensated metals 
for which we have to take into account the multiband conduction electrons.  
Nevertheless, enhancement occurs in the first and second terms in 
eq.(\ref{eq:8a}) in proportion to 
$-\partial\ln N^{*}(\epsilon)/\partial\epsilon|_{\epsilon=\mu}$ in each 
heavy fermion band, maintaining the validity of the result for a single-band 
Fermi liquid.  
This is because the Fermi level is located just below the level corresponding 
to the divergence of DOS due to the hybridization effect, which causes the 
hybridization gap in the case of a single conduction band.  A crucial point 
is that the sign of $\partial\ln N^{*}(\epsilon)/\partial\epsilon$ and 
the inverse mass tensor do not depend on 
the sign of the velocity but on the curvature of the dispersion of 
quasiparticles.  This explains why the $q$ value is positive for 
Sr$_2$RuO$_4$ even though the $\gamma$-band (heaviest band) is electron-like 
with the negative Hall coefficient, $R_{\rm H}<0$.  Indeed, the energy 
derivative of the DOS of the $\gamma$-band is positive due to the van Hove 
singularity located just above the Fermi level~\cite{Mazin}.  It is also 
the case in (BEDT-TTF) salt reported in ref.\ 5.  Indeed, the 
DOS in the H\"uckel type tight-binding theory seems to explain the difference 
in the sign of the Seebeck coefficient between the $b$- and 
$c$-directions~\cite{Mori}.  

A prediction based on the above result is that CeRu$_2$Si$_2$ with $q\sim+1$ 
without the magnetic field $H=0$ changes the sign of $q$ just above the 
metamagnetic field $H_{\rm M}$ for the temperature gradient along the 
$c$-axis.  This is because the singular peak of the 
DOS arising from the flat band structure around $(0,0,\pi/c)$, which is 
located above the Fermi level~\cite{Aoki,IkedaThesis}, is expected to shift 
down to the Fermi level at $H>H_{\rm M}$, leading to the reversal of the 
signs of $\partial\ln N^{*}(\epsilon)/\partial\epsilon$ and the inverse mass 
tensor.  

\subsection*{$f^{2}$-Based Heavy Fermions}
The above theory for heavy fermions is valid for $f^{1}$-based compounds.  
On the other hand, $f^{2}$-based heavy fermions, such as U- and Pr-based 
compounds, have a different structure of the quasiparticle band.  
In such systems, one has to take into account the plural $f$-orbitals split by 
the crystalline electric field (CEF) effect.  It gradually became apparent 
in the mid 90s that the effective CEF splitting is considerably suppressed 
to be less than the renormalized energy scale, the effective Fermi energy, 
while that of the $f^{1}$-based systems is slightly enhanced due to the 
correlation effect~\cite{Trees,Ikeda}.  Indeed, this is the only way for the 
mass of quasiparticles to be highly enhanced by satisfying the 
two conditions, the Landau-Luttinger sum rule~\cite{Luttinger,AGD} 
and the requirement that the 
$f$-electron number $n_{i\rm f}$ of the $i$-th orbital ($i=1,2$) at each 
site be nearly unity, $n_{i\rm f}\simeq 1$, as seen below.  It was shown 
on the basis of the investigation of the $f^{0}$-$f^{1}$-$f^{2}$ 
model~\cite{Ikeda} that the mass enhancement arises only if the $f$-electron 
number is nearly 1 or 2, as long as the two $f$-orbitals with a low-lying CEF 
level are relevant.  

\begin{figure}[h]
\begin{center}
\rotatebox{0}{\includegraphics[width=0.8\linewidth]{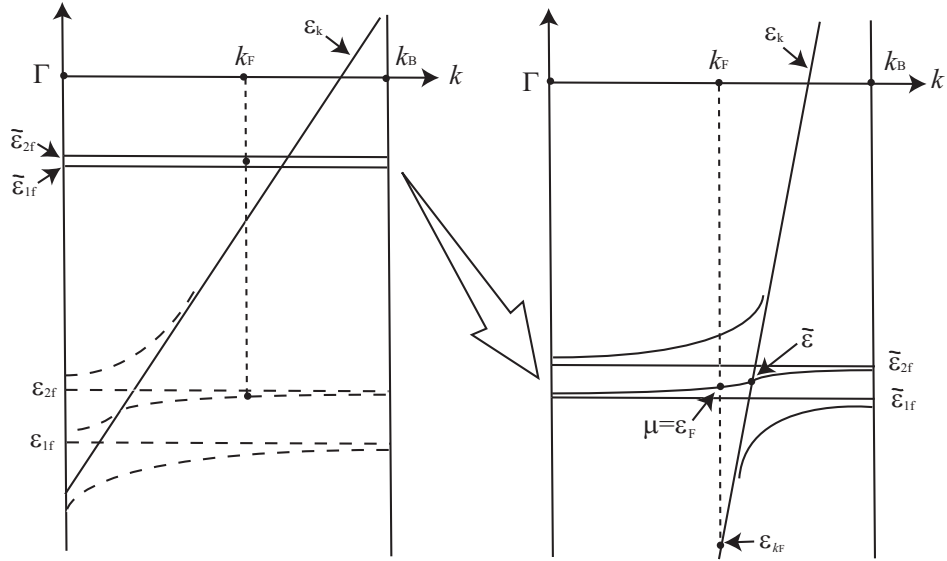}}
\caption{
Schematic view of the dispersion of $f^{2}$-based heavy fermions in some 
direction in the Brillouin zone.  The dispersion $\epsilon_{k}$ of conduction 
electrons is assumed to be linear, and $k_{\rm F}$ and $k_{\rm B}$ denote 
the Fermi wavevector and the zone boundary, respectively.  
The dashed curves are for a noninteracting system, and the solid curves are 
for a renormalized dispersion.  The energy scale of the right-hand panel is 
enlarged compared to the left-hand one.  It is noted that 
effective CEF splitting is reduced to be smaller than the characteristic 
energy scale $T_{0}$.  
}
\label{Fig:2}
\end{center}
\end{figure}

The effective Hamiltonian for the quasiparticles of $f^{2}$-based heavy 
fermions is given in the form:
\begin{eqnarray}
H_{\rm qp}&=&\sum_{{\vec k}}\Bigg\{\epsilon_{{\vec k}}
c^{+}_{{\vec k}\sigma}c_{{\vec k}\sigma}+
\sum_{i=1,2}
{\tilde \varepsilon}_{i\rm f}f^{+}_{{i\vec k}\sigma}f_{i{\vec k}\sigma}
\nonumber
\\
& &\qquad+\sum_{i=1,2}
\left[{\tilde V}_{i{\vec k}}c^{+}_{i{\vec k}\sigma}f_{i{\vec k}}
+{\rm h.c.}\right]\Bigg\},
\label{eq:26}
\end{eqnarray}
where the renormalized $f$-levels ${\tilde \varepsilon}_{i{\rm f}}$'s and 
the renormalized hybridizations ${\tilde V}_{i{\vec k}}$'s are given by the 
orbital-dependent $f$-electron self-energy and the renormalization amplitude 
as 
\begin{equation}
{\tilde \varepsilon}_{i\rm f}\equiv
a_{i\rm f}[\varepsilon_{i\rm f}+\Sigma_{i\rm f}(\mu)],
\label{eq:27}
\end{equation} 
and ${\tilde V}_{i{\vec k}}\equiv\sqrt{a_{i\rm f}}V_{i{\vec k}}$.  
The dispersion of the quasiparticles $E_{k}$ is given as three solutions of 
\begin{equation}
\prod_{i=1,2}({\tilde \varepsilon}_{i\rm f}-E_{k})(\epsilon_{\vec k}-E_{k})=
\sum_{i=1,2}{\tilde V}_{i}^{2}({\tilde \varepsilon}_{i\rm f}-E_{k}).
\label{eq:28}
\end{equation}
The schematic behavior of the dispersion is shown in Fig.\ \ref{Fig:2}.  
The renormalized $f$-levels are lifted from their original positions 
in such a way that the total electrons per site are distributed to two 
$f$-electrons and the rest to the conduction electrons.  The number of 
correlated electrons is counted using the Landau-Luttinger sum rule, while 
that of the conduction electrons is counted by summing up the $k$-points 
below the Fermi energy.  
Then, it is easily seen that the two $f$-electrons occupy the orbital of 
the lower level ${\tilde \varepsilon}_{1\rm f}$ unless the renormalized level 
splitting ${\tilde \Delta}\equiv
{\tilde \varepsilon}_{2\rm f}-{\tilde \varepsilon}_{1\rm f}$ is far less than 
the width of each band $\sim a_{\rm f}|{\tilde V}|^{2}/D$.  
On the other hand, if ${\tilde \Delta}$ is significantly smaller than 
$a_{\rm f}|{\tilde V}|^{2}/D$, it is possible to satisfy the above two 
conditions because the lower level then consists of both $f$-orbitals with 
nearly equal weight.  This is the reason why 
the renormalized level splitting is considerably reduced to make the 
$f^{2}$-based heavy fermion band~\cite{Trees,Ikeda}.  It is not difficult 
to see that eq.(\ref{eq:28}) has a trivial solution at some $k$ as 
\begin{equation}
E_{k}=\epsilon_{k}={\tilde \varepsilon}
\equiv{{\tilde V}_{1}^{2}{\tilde \varepsilon}_{1\rm f}+
{\tilde V}_{2}^{2}{\tilde \varepsilon}_{2\rm f}\over
{\tilde V}_{+}^{2}},
\label{eq:29}
\end{equation}
where ${\tilde V}_{\pm}^{2}\equiv({\tilde V}_{1}^{2}\pm{\tilde V}_{2}^{2})$.  

Formal solutions of eq.(\ref{eq:28}) near the renormalized $f$-levels 
are given similarly to eq.(\ref{eq:13}) as 
\begin{equation}
E_{k}^{\pm}={\tilde \varepsilon}
+{{\tilde V}_{-}^{2}\over2{\tilde V}_{+}^{2}}{\tilde \Delta}
-{{\tilde V}_{+}^{2}\over2{\bar \epsilon}_{k}}
\pm\sqrt{{{\tilde \Delta}^{2}\over4}+
\left({{\tilde V}_{+}^{2}\over2{\bar \epsilon}_{k}}\right)^{2}-
{{\tilde V}_{-}^{2}\over2{\bar \epsilon}_{k}}{\tilde \Delta}}\quad,
\label{eq:30}
\end{equation}
where ${\bar \epsilon}_{k}\equiv\epsilon_{k}-E^{\pm}_{k}$ and 
$|{\bar \epsilon}_{k}|\sim D$ except for the narrow region 
$|\epsilon_{k}-{\tilde \varepsilon}|<{\rm max}
({\tilde \Delta},{\tilde V}_{+}^{2}/D)$.  
In the situation shown in Fig.\ \ref{Fig:2}, the Fermi level is located in 
the band $E^{-}$ with $\epsilon_{k_{\rm F}}<{\tilde \varepsilon}$, which 
simplifies the following analysis.  
In order to avoid the double occupancy of the orbital-1 with a lower CEF 
energy, $\epsilon_{k_{\rm F}}$, the energy of 
the conduction band corresponding to the Fermi wave number should be slightly 
lower than ${\tilde \varepsilon}$ so that 
$|{\bar \epsilon}_{k_{\rm F}}|\gg|{\tilde \epsilon}_{\rm f}-E^{-}_{k}|$, 
$|{\tilde V}_{+}|$, because the weight in the lower level 
${\tilde \varepsilon}_{1\rm f}$ of the orbital-1 is slightly larger than 
that of orbital-2.  

The DOS near the Fermi level is given as 
\begin{equation}
N^{*}(\epsilon)\simeq 4N^{\rm cond}
{(\epsilon_{k}-\epsilon)^{2}\over{\tilde V}_{+}^{2}}
\biggl|{{\displaystyle \epsilon-{\tilde \varepsilon}+{\Delta_{0}\over 2}
+{{\tilde V}_{+}^{2}\over 2(\epsilon_{k}-E^{-}_{k})}}
\over\epsilon-{\tilde \varepsilon}}\biggr|,
\label{eq:32}
\end{equation}
where $\Delta_{0}\equiv{\tilde \Delta}{\tilde V}_{-}^{2}/{\tilde V}_{+}^{2}$, 
and 
the term of ${\cal O}({\tilde V}_{+}^{2}/(\epsilon_{k}-E^{-}_{k})^{2})$ has 
been neglected.  
Then, the logarithmic derivative of DOS at the Fermi level is expressed as 
\begin{equation}
{\partial\over\partial\epsilon}\ln N^{*}(\epsilon)\biggl|_{\epsilon=\mu}
\simeq
-{4\over T_{0}}\times
{{\displaystyle 
({\tilde \varepsilon}-\mu)^{2}+({\tilde \varepsilon}-\mu)(T_{0}-\Delta_{0})
+{1\over 4}(T_{0}-\Delta_{0})\left({T_{0}\over 2}-\Delta_{0}\right)
}\over{\displaystyle 
({\tilde \varepsilon}-\mu)
\left({\tilde \varepsilon}-\mu+{T_{0}-\Delta_{0}\over 2}\right)
}}
\equiv-{4J\over T_{0}}<0,
\label{eq:33}
\end{equation}
where the characteristric energy scale $T_{0}$ is defined as 
$T_{0}\equiv{\tilde V}^{2}_{+}/|{\bar \epsilon}_{k_{\rm F}}|$.  
The inverse mass tensor at the Fermi level is given as 
\begin{eqnarray}
{\partial^{2} E^{-}_{k}\over\partial k_{\alpha}^{2}}\biggl|_{\epsilon=\mu}
&\simeq&{T_{0}\over|{\bar \epsilon}_{\rm k_{\rm F}}|^{2}}\times
{{\tilde \varepsilon}-\mu\over {\displaystyle {\tilde \varepsilon}-\mu+
{T_{0}-\Delta_{0}\over 2}}}\times
\nonumber
\\
& &\biggl[1+{T_{0}(T_{0}-\Delta_{0})\over
{\displaystyle 8\left({\tilde \varepsilon}-\mu+
{T_{0}-\Delta_{0}\over 2}\right)^{2}}}\biggr]
\left({\partial\epsilon_{k}\over\partial k_{\alpha}}\right)^{2}.
\label{eq:34}
\end{eqnarray}
With the use of expression (\ref{eq:34}) and a similar expression for 
the velocity $\partial E_{k}^{-}/\partial k_{\alpha}$, 
the second term in the brace of eq.(\ref{eq:8a}) is expressd as 
\begin{equation}
{\sum_{{\vec k}}\delta(\epsilon-E_{{\vec k}})
\partial v_{\alpha}({\vec k})/\partial k_{\alpha}\over
\sum_{{\vec k}}\delta(\epsilon-E_{{\vec k}})
v_{\alpha}({\vec k})v_{\alpha}({\vec k})}\biggr|_{\epsilon=\mu}
={4\over T_{0}}\times{
{\displaystyle \left(
{\tilde \epsilon}-\mu+{T_{0}-\Delta_{0}\over 2}\right)^{2}+
{1\over 8}T_{0}(T_{0}-\Delta_{0})}\over
{\displaystyle 
({\tilde \epsilon}-\mu)
\left({\tilde \epsilon}-\mu+{T_{0}-\Delta_{0}\over 2}\right)
}}
\equiv{4H\over T_{0}}>0.
\label{eq:35}
\end{equation}

Thus, with the use of expressions (\ref{eq:33}) and (\ref{eq:35}) , 
the factor ${\cal F}$ in eq.(\ref{eq:8a}) is expressed as 
\begin{equation}
{\cal F}={4\over T_{0}}\times{H+J+(H-J)[\pi z(\mu)N^{*}(\mu)u]^{2}
\over 1+[\pi z(\mu)N^{*}(\mu)u]^{2}}>0.
\label{eq:36}
\end{equation}
Since $H$ and $J$ are positive constants of ${\cal O}(1)$ and 
$N^{*}(\mu)T_{0}\sim1$, the quasi-universal value of $q\sim -1$ follows as 
in the argument for the single-band Fermi liquid.  This result explains the 
fact that $q\sim -1$ for UBe$_{13}$ and URu$_2$Si$_2$~\cite{Behnia}.  
A straightforward prediction is that the Pr-based heavy fermions with the 
filled Skutterudite structure, such as PrFe$_4$P$_{12}$ and PrOs$_4$Sb$_{12}$, 
will exhibit the universal ratio $q\sim-1$~\cite{SatoH}.  

In contrast, UPd$_2$Al$_3$ with $q\sim +1$ is considered to 
be in the $f^{3}$-configuration~\cite{Sato} and the situation would be 
the same as in the Ce-based compounds, which is consistent with $q\sim +1$.  
This is because three $f$-electrons exhibit itinerant-localized dual 
behavior: two of them are localized to form an $f^{2}$ CEF level structure and 
one of them forms the quasiparticles~\cite{Yotsuhashi}.

We acknowledge K. Behnia for directing our attention to the present problem, 
and stimulating discussions.  One of us (K.M.) benefitted from a lively 
conversation with D. Jaccard and J. Flouquet on clarifying the implications 
of the results, and information on PrFe$_4$P$_{12}$ from H. Sato and 
a FLEX calculation from K. Hoshihara.  
This work is supported by a Grant-in-Aid for Scientific Research in Priority 
Areas (No.16037209) from Monbu-Kagakusho, and a Grant-in-Aid for Scientific 
Research (No.16340103) and the 21st Century COE Program from the Japan Society 
for the Promotion of Science.


\begin{thebibliography}{99}
\bibitem{Phonon2}
In electron-phonon system, 
$\lim_{\epsilon\to\mu}\partial \ln z(\epsilon)/\partial\epsilon
=\lim_{\epsilon\to\mu}(z^{-1}-1)\partial \ln N^{*}(\epsilon)
/\partial\epsilon$, 
as discussed in the paper of ref.\ 9, so that the value $q$ is affected to 
some extent.  Nevertheless, it does not alter the quasi-universality of 
$q$ in the logarithmic scale because $(z^{-1}-1)\sim{\cal O}(1)$.
\bibitem{comment2}
The mass enhancement, proportional to 
$\{{\rm const.}-[(\epsilon-\mu)^{2}+\omega_{0}^{2}]^{1/4}\}$, is obtained 
by taking average of the mass tensor 
$1/|\partial v_{\alpha}/\partial k_{\alpha}|$, which diverges logarithmically 
on the hot lines, over the Fermi surface.
\end{thebibliography}
\end{document}